\documentstyle[12pt, amssymb]{article}

\def\fnote#1#2{\begingroup\def\thefootnote{#1}\footnote{#2}\addtocounter{footnote}{-1}\endgroup}

\def\beq{\begin{equation}}
\def\eeq{\end{equation}}
\def\bea{\begin{eqnarray}}
\def\eea{\end{eqnarray}}
\def\lleq#1{\label{#1}\eeq}

\def\lra{\longrightarrow}

\def\lra{{\longrightarrow}}

\def\vphi{{\varphi}}

\def\rmA{{\rm A}}
\def\rmB{{\rm B}}

\def\rmdim{{\rm dim}}   \def\rmlcm{{\rm lcm}}  
\def\rmwt{{\rm wt}}

\def\rmBP{{\rm BP}}
   \def\rmGal{{\rm Gal}}  \def\rmND{{\rm ND}}   \def\rmSpec{{\rm Spec}}

   \def\si{{\sigma}}
 \def\G{{\Gamma}}
 \def\tPhi{{\tilde{\Phi}}}   \def\tPsi{{\tilde{\Psi}}}
 \def\wtPhi{{\widetilde{\Phi}}}  \def\wtPsi{{\widetilde{\Psi}}}

 \def\oell{{\overline{\ell}}}  \def\oq{{\overline{q}}} \def\os{{\overline{s}}}

 \def\ochi{{\overline{\chi}}}

   \def\cI{{\cal I}}

\def\mathC{{\mathbb C}}
\def\mathN{{\mathbb N}}  \def\mathP{{\mathbb P}}  \def\mathQ{{\mathbb Q}}  \def\mathZ{{\mathbb Z}}

\def\Om{{\Omega}}

\def\a{\alpha}   \def\b{\beta}      
 \def\G{\Gamma}     
    \def\Om{\Omega} \def\si{\sigma}
\def\Si{\Sigma}       

 \def\mathC{{\mathbb C}} \def\mathF{{\mathbb F}}
  \def\mathN{{\mathbb N}}
\def\mathP{{\mathbb P}}
\def\mathQ{{\mathbb Q}} 
\def\mathZ{{\mathbb Z}}

 \def\wtPhi{{\widetilde{\Phi}}}  \def\wtPsi{{\widetilde{\Psi}}}

\def\rmA{{\rm A}}  \def\rmB{{\rm B}}

            \def\rmBP{{\rm BP}}

 \def\rmGal{{\rm Gal}}           \def\rmGSO{{\rm GSO}}

\def\rmND{{\rm ND}}

            \def\rmSpec{{\rm Spec}}

\def\rmdeg{{\rm deg}}           
\def\rmdim{{\rm dim}}

\def\rmlcm{{\rm lcm}}

     \def\rmmod{{\rm mod}}

\def\rmth{{\rm th}}         
    
   \def\rmwt{{\rm wt}}

\def\oell{{\overline{\ell}}}  \def\oq{{\overline{q}}} \def\os{{\overline{s}}}

\def\ochi{{\overline{\chi}}}

\def\beq{\begin{equation}}
\def\eeq{\end{equation}}
\def\bea{\begin{eqnarray}}
\def\eea{\end{eqnarray}}
\def\llea#1{\label{#1}\eea}
\def\lleq#1{\label{#1}\eeq}

\let\nn=\nonumber

\begin{document}

\phantom{\hfill \today}

\vskip .6truein

 \centerline{\Large {\bf Motivic $L$-Function Identities from CFT}}

 \vskip .1truein

 \centerline{\Large {\bf and Arithmetic Mirror Symmetry}}

 \vskip .5truein

\centerline{\sc Rolf Schimmrigk\fnote{\diamond}{netahu@yahoo.com; rschimmr@iusb.edu}}

\vskip .2truein

\centerline{Indiana University South Bend}

\centerline{1700 Mishawaka Ave., South Bend, IN 46634}

\vskip .9truein

\baselineskip=19pt

\centerline{\bf Abstract.}

\begin{quote}
  Exactly solvable mirror pairs of Calabi-Yau threefolds of hypersurface type
  exist in the class of Gepner models that include nondiagonal affine invariants.
  Motivated by the string theoretic automorphy established previously for models
  in this class it is natural to ask whether the arithmetic structure of mirror pairs varieties
  reflects the fact that as conformal field theories they are isomorphic.
  Mirror symmetry in particular predicts that the $L$-functions of the
  $\Om$-motive of such pairs are identical.
  In the present paper this prediction is confirmed by showing that the $\Om$-motives
  of exactly solvable mirror pairs are isomorphic. This follows as a corollary of the proof of
  a more general result establishing an isomorphism between nondiagonally and diagonally
  induced motives in this class of varieties.
   The motivic approach formulated here circumvents the difficulty that no
  mirror construction of the Hasse-Weil zeta function is known.

\end{quote}

\renewcommand\thepage{}
\newpage

\pagenumbering{arabic}

\baselineskip=16pt

\tableofcontents

\baselineskip=21.6pt 
\parskip=.15truein
\parindent=0pt

\vskip .4truein

\section{Introduction}

The purpose of this paper is to establish an isomorphism between motives associated to
 certain types of hypersurfaces that arise in the context of exactly solvable string
 compactifications.
The motivation for this result comes from two independent, but related, programs.
 The first is to understand the relation between the  conformal field theory $T_\Si$
 on the string worldsheet $\Si$  and the structure of Calabi-Yau varieties $X$ by relating
 automorphic forms associated to motives $M(X)$ to  automorphic forms derived from
 Kac-Moody algebras.  The second aims at an understanding of  mirror symmetry by using
methods from arithmetic geometry.

Arithmetic mirror symmetry is an old issue that  is made difficult by the fact that the
 mirror transformation exchanges even and odd cohomology of the variety. This implies that the
 basic arithmetic object associated
to the variety $X$, given by the local zeta functions $Z(X/\mathF_p,t)$  defined for the finite fields $\mathF_p$
 for any prime $p \in \mathN$, does not respect mirror symmetry, a fact that follows from the factorization
of the zeta function into cohomological pieces, envisioned by Weil \cite{w49} and proven by
 Grothendieck \cite{g65}.
 Hence the arithmetic approach appears to suffer from the usual problems of understanding
mirror symmetry in the geometric setting (see e.g. \cite{cdov04} for a more detailed
 discussion of this issue).

Guidance in this problem is provided by results which establish that the $L$-functions of
 certain varieties are identical to $L$-functions determined by modular
forms on the string worldsheet (see e.g. \cite{rs08} and references
therein). Such an identity implies that the mirror $L$-function must
be identical to the original one because the underlying conformal
field theories of the varieties are isomorphic, hence the modular
forms on the worldsheet are identical. This prediction of mirror
symmetry was confirmed in the context of models with central charge
 $c=3$ in ref. \cite{rs07} and for some rigid mirror pairs in \cite{kls08}.
 The advantage of elliptic curves is that their $L$-functions
 can be obtained directly by counting
 the points of $X/\mathF_p$ on the variety, while the cases of rigid mirrors only
  involved weighted Fermat hypersurfaces. In the more general case the extension of
this test is made more difficult because it is necessary to
have a definition of motives for nondiagonal varieties.  Such a construction has recently been given
in \cite{rs13b} and this result will be applied here in the context of exactly solvable string models
to establish an isomorphism of their $\Om$-motives, and to discuss its implications,
 in particular in the context of mirror symmetry. This isomorphism in particular implies the identity
 of the $L$-functions of the $\Om$-motives of mirror pairs even though the Hasse-Weil zeta function
is not invariant under mirror symmetry.

The outline of this article is as follows. Section 2 introduces the map from diagonal to
 nondiagonal varieties of $D$-type. Section 3 defines the relevant motives and
 proves the motivic isomorphism. Section 4 considers applications of the isomorphism
 in the context of the automorphic spacetime program and mirror symmetry, while the final section
 briefly discusses a second map from diagonal to nondiagonal manifolds.

\section{The $AD$-space of hypersurfaces}

The focus in the following is on certain types of  hypersurfaces  $X_n$
 of dimension $n$ embedded in weighted projective spaces $\mathP_{(w_0,...,w_{n+1})}$ with
 weights $w_i \in \mathN$. The specific class of interest is generated by monomial and
 nondiagonal binomial building blocks given by
 \bea
 p_A^{(i)} (z_i) &= & z_i^{d_i}  \nn \\
 p_D^{i)}(z_i,z_{i+1}) &=& z_i^{d_i}+ z_iz_{i+1}^{d_i+1}.
\eea
 The space of polynomials generated by $p_A$ and $p_D$ over a field $K$ then consists of varieties
 of the type
\begin{equation}
  X_n ~=~ \left\{\sum_i \a_i p_A^{(i)} + \sum_j \b_j p_D^{(j)}
   ~=~ 0 ~{\Big |}~ \a_i,\b_j \in K\right\} ~\subset~ \mathP_{(w_0,...,w_{n+1})},
 \end{equation}
 with $n=n(i,j)$ depending on the range of $i$ and $j$, and is denoted by
 \beq
  H_K ~=~ \langle p^{(i)}_A, ~p^{(i)}_D\rangle_K.
 \eeq

On the space $H_K$ one can define a map that increases the number of links by replacing monomials
by  nondiagonal binomials.  More precisely, consider any hypersurface $X$ defined by a
polynomial containing at least one  purely monomial term and a "trivial factor", i.e. it is of
  type
 \beq
 p_A(z_i,z_{i+1}) := z_i^a + z_{i+1}^2,~~a\in 2\mathN,
  \eeq
 and denote by $X_D$ the variety obtain by replacing $p_A$ by the nondiagonal factor
  \beq
 p_D(z_i,z_{i+1}) := z_i^{a/2} + z_iz_{i+1}^2.
  \eeq
 The replacement $p_A(z_i,z_{i+1}) \longmapsto p_D(z_i,z_{i+1})$  induces a map on the space of all
 CY hypersurfaces
 \beq
 s_D: ~ H_K ~\lra ~ H_K,
  \eeq
 where $s_D$ is extended to the whole space $H_K$ by defining it to be the identity on those 
  polynomials
 that do not contain a summand of the appropriate form.

\section{$A$- and $D$-motives}

The image $s_D(X)$ of a variety $X$ is in general topologically different from $X$, indicated e.g.
 by their different Hodge numbers. The  fact that the $L$-function of the $\Om$-motive
 of diagonal varieties has been shown in several instances to be determined by the modular forms
 defined by the conformal field
 theory $T_\Si$ on the worldsheet $\Si$ raises the question whether the $\Om$-motive
 of both varieties is the same even though the manifolds are topologically distinct.
 The purpose of this section is to establish such an isomorphism.

\subsection{Diagonal and nondiagonal $\Om$-motives}

The notion of the $\Om$-motive is universal, applicable to all Calabi-Yau varieties and Fano varieties
 of special type \cite{rs08}. For the hypersurfaces considered in the present paper this general
construction can be made explicit and computable.  For diagonal  varieties
this has been described in the above reference.
For nondiagonal varieties the construction of the $\Om$-motives needs
 to be generalized. This is briefly reviewed below and more details can be found in
 \cite{rs13b}. Denote by $X^d_n$ any diagonal hypersurface of type $A$ or $D$
 of complex dimension $n$ and degree $d$
 embedded in a weighted projective space.  The polynomial defining the hypersurface contains monomials
 of the variables $z_i$ occurring with degrees $d_i$. For hypersurfaces of tadpole type defined by
 polynomials of the type 
  $p=\sum_{i=0}^n z_i^{d_i} + z_nz_{n+1}^{d_{n+1}}$ define the integer
  $v=\rmlcm\{d_i\}_{i\neq n} \in \mathN$ and consider
  the $v^\rmth$ root of unity $\xi_v=e^{2\pi i/v}$
 and the cyclic group $\mu_v = \langle \xi_v\rangle$ it generates. Associate to $X=X_n^d$
 the abelian number field
  $K_X = \mathQ(\mu_v)$ and denote its Galois group by $\rmGal(K_X/\mathQ)$.
 The cohomological realization $H(M_\Om)$ of the
 $\Om$ motive $M_\Om$ now is defined as the Galois orbit defined by
 the action of $\rmGal(K_X/\mathQ)$ on the holomorphic $n-$form \cite{rs13b}.
For diagonal varieties the integer $v$ is given by the degree of the hypersurface.

For varieties of the type considered here the definition just given can be
made more explicit because
 the cohomology class $\Om$ can be represented as a unit vector
 $u_\Om=(1,...,1) \in \mathZ^{n+2}$
 and the action of the Galois group on $\Om$ can be expressed as the action of
 $\rmGal(K_X/\mathQ) \cong (\mathZ/v\mathZ)^\times$ on this vector. The actions differ for the
diagonal and the nondiagonal cases. For diagonal hypersurfaces
 the action of $\si_r \in \rmGal(K_d/\mathQ)$
is given by $\si_r(u_\Om^i) \equiv ru_\Om^i(\rmmod~d_i)$ for all $i=0,...,n+1$, while in the
 nondiagonal case the action of $\si_r \in \rmGal(K_v/\mathQ)$
is given by a combination of an analogous modding relation in combination with
 the fact  that the resulting images $u_r$ of $u_\Om$ under $\si_r$ 
 have to satisfy the constraint
 $d|\sum_i u_r^iw_i$ \cite{rs13b}.
 With these definitions the  cohomological realization of
the  $\Omega$-motive $M_\Omega$  of a hypersurface can be written as
 \beq
  H(M_\Om) ~=~ \langle \rmGal(K_v(X)/\mathQ, ~ \Om\rangle
 \eeq
 where $\Om$ is represented by $u_\Om$.

\subsection{The motivic $AD$-isomorphism}

 Let $X_D$ be a variety  generated by the action of $s_D$ on a variety $X$. Then the
 $\Om$-motive $M_\Om(X)$ of $X$ is isomorphic to the
 $\Om$-motive $M_\Om(X_D)$ of $X_D=s_D(X)$. As a consequence of this
 motivic $AD$-isomorphism
 the $L$-function of the $\Omega$-motive of $X_D$ is the same as the $L$-function of the
 $\Omega$-motive of  $X$
  \begin{equation}
    L_\Omega(X_D,s) ~=~ L_\Omega(X,s).
\end{equation}

 This result will be derived in the remainder of the section
from the structure of the underlying $N=2$ superconformal
 field theory.  Applications  to motives of Calabi-Yau varieties
and to mirror symmetry will be indicated in the following section.

\subsection{Ingredients of the proof}

The basic idea is to argue conformal field theoretically.  The proof given below uses the fact
 that the cohomology
 of certain types of Calabi-Yau hypersurfaces is determined by the massless spectrum
 of tensor products of $N=2$ superconformal minimal models endowed with a
projection that ensure integral U(1) charges \cite{g88}.

\subsubsection{Exactly solvable tensor products}

  Minimal models with two supersymmetries are rational conformal field
 theories constructed from the affine algebra $A_1^{(1)}$ at conformal
levels $k$ with central charge
  \begin{equation}
  c=\frac{3k}{k+2},~~~~~ k \in \mathN.
  \end{equation}
 Such models are denoted by $k\left(A_1^{(1)}\right)$, with notation
 $k_A$ and $K_D$ if the affine
invariant needs to be indicated.
 In order for an $N=2$ superconformal theory to describe a variety $X$
 the central charge has to satisfy the dimension constraint
  \beq
   c ~~=~ 3\rmdim_\mathC X,
  \eeq
  hence minimal models need to be tensored in order to saturate the central charge constraint
  \begin{equation}
  \bigotimes_{i=0}^{n+1} k_i\left(A_1^{(1)}\right): ~~\phantom{wh}
   c ~=~ \sum_{i=0}^{n+1} \frac{3k_i}{k_i+2}.
 \label{central-charge}\end{equation}

There are two ingredients in the proof of the $L$-function relation for varieties
associated to a tensor product $\otimes_i k_i(A_1^{(1)})$.
 In order to formulate them it is necessary to review briefly the relevant structures of such theories.
 In tensor products these fields are obtained from the fields of the individual factors
 $\vphi = \prod_i \vphi_i$, where $\vphi_i \in \rmSpec(k_i(A_1^{(1)}))$ are fields in the
 individual factors.  Since $N=2$ superconformal minimal models are rational
 there are a finite number of fields and  the parametrization can be obtained explicitly in
terms of quantum numbers given by $(\ell,q,s)$, denoted here by
  $\varphi^{k}_{\ell,q,s} \in \rmSpec(k(A_1^{(1)}))$,
 where the ranges of the quantum numbers $\ell,q,s$ are
 determined by the level $k$.  The conformal weight and U(1) charges of these
fields are given by
 \bea
 \Delta^k_{\ell,q,s} &=& \frac{\ell(\ell+2)-q^2}{4(k+2)} + \frac{s^2}{8}
 \nonumber \\
 Q^k_{\ell,q,s} &=& - \frac{q}{k+2} + \frac{s}{2},
 \eea
 with $\ell\in \{0,1,...,k\}$, $\ell+q+s\in 2\mathZ$, and $|q-s|\leq \ell$ \cite{g88} and
the identification $(\ell,q,s) \cong (k-\ell,q+k+2,s+2)$.
 The fields which correspond to the cohomology of the variety are determined by the subset  of
chiral primary fields of the theories,
 characterized by the condition that their anomalous dimensions
 $\Delta$ and their U(1) charge $Q$ are related as $Q(\phi) = 2\Delta(\phi)$. Combining these building
 blocks into marginal operators with specific integrality constraints for the total charge
 accounts for the cohomology
 of the critical theory.

\subsubsection{Modular invariants}

The detailed spectrum of the theory
depends on the modular invariants $N_{\ell,\oell}, M_{q,\oq}, R_{s\os}$ that appear in the partition
function
  \beq
   Z^{k} ~=~ \sum N_{\ell, \oell} M_{q,\oq} R_{s,\os} \chi^{k}_{\ell,q,s} \ochi^{k}_{\oell,\oq,\os},
  \eeq
 where $\chi^k_{\ell,q,s}$ denote the basic characters. Most important in the present context is the
 modular invariant $N_{\ell,\oell}$ associated to the affine part of the $N=2$ superconformal theory.
 These invariants have been classified and follow an ADE pattern \cite{ciz87a, ciz87b}. It is known in particular
that for each even conformal level $k\in 2\mathN$ the partition function is modular invariant
with both the diagonal $A$-invariant as well as the $D$-invariant.

The choice of the modular invariants therefore determines how the left- and right-moving sectors are
tensored together. In the following these tensor product fields will be denoted by
 $\phi_{\ell,q,s}^{\oell,\oq,\os}$, neglecting the level index $k$. Here the $(\ell,q,s)$
 denotes the left-moving sector, while $(\oell,\oq,\os)$ denotes the right-moving sector.

\subsection{The proof}

  The basic observation in the proof of the motivic $AD$-isomorphism is that
the map $s_D$ is the geometric counterpart of the replacement of the diagonal
$A$-invariant in the partition function of a level $k-$minimal model by an affine $D$-invariant
  \beq
  A_1^{(1)}: ~~~ k_A ~\lra~ k_D,
  \eeq
  where the affine invariants of the
  remaining factors of the tensor product $\otimes_i k_i$  are left unchanged.
 There are two steps in the proof that the motives are isomorphic, with a resulting
  $L$-function identity. These two steps reflect the definition of the $\Om$-motive outlined above:
 \parskip=0pt
 \begin{enumerate}
   \item The first ingredient is that the field $\vphi_\Om$
  corresponding to the holomorphic form $\Om$ is
  unchanged in the transition $k_A\longmapsto k_D$.

 \item The second ingredient is that not only is $\vphi_\Om$ remain unchanged, but that
  the Galois group also remains the same.

 \end{enumerate}

 For nondiagonal hypersurfaces the motive is determined by the orbit of the
   Galois group of $\mathQ(\mu_v)$, where
 $v=\rmlcm\{d_i\}_{i\neq n}$, while for 
$A$-motives the rank of the motives is determined by the degree of the
variety $d=\rmdeg(X)$.

\parskip=0.15truein

\subsubsection{Part I: invariance of the holomorphic form $\Om$}

The basic reason for the invariance of $\vphi_\Om$ under the map $s_D$ is because in the transition
  all factors except one  remain the same. It follows that
 the factor $\vphi_{\ell_A,m_A,s_A}^{k_A}$ that is changed must be
replaced by a chiral primary field of exactly the same weight and the same charge, which will imply
 that the field must be invariant under the exchange of the affine invariant.

It is useful to first recall that the $D$-theory is the resolution of the $\mathZ_2$-quotient
 of the diagonal model at level $k$
 \beq
  k_D ~=~ {\rm res}(k_A/\mathZ_2),
 \eeq
 where $\mathZ_2$ is short-hand notation for $\mathZ/2\mathZ$.  This quotient isomorphism
 has been discussed in detail  in \cite{ls90}  in  Landau-Ginzburg framework of
 \cite{m88,vw88,lvw89,w93}.
 The resolution of the
 quotient theory introduces a single twist field, denoted by $\tPsi$, and
the two models can described by the superpotentials
 \bea
 W_A &=& \Phi^{k+2} + \Psi^2 \nn \\
 W_D &=& \tPhi^{\frac{k+2}{2}} + \tPhi \tPsi^2,
 \llea{superpotentials}
 with a map
 \beq
 \Phi ~=~ \wtPhi^{1/2}, ~~~~\Psi ~=~ \wtPhi^{1/2}\Psi,
 \lleq{fractional-transf}
 involving fractional exponents, hence called fractional transform in \cite{ls90, ls95}. The ideal
 of the $D$-theory is generated by
 \beq
  \cI_D ~=~ \left\langle \wtPhi^{\frac{k}{2}} + \wtPsi^2, ~\wtPhi\wtPsi \right \rangle
 \eeq
 leads to the spectrum
 \beq
 \rmSpec(k_D) ~=~ \{1,\wtPhi,...,\wtPhi^{\frac{k}{2}}, ~\wtPsi\}.
  \eeq
 The weights of the fields in the $D$-model from the superpotential as
  \beq
 \rmwt(\wtPhi) ~=~ 2\rmwt(\Phi) ~=~ \frac{2}{k+2},  ~~~~~~~~~
 \rmwt(\wtPsi) ~=~ \frac{k}{2(k+2)},
 \eeq
 consistent with the structure of the ideal.

In order to make the CFT theoretic realization of the holomorphic $\Om$ form explicit
 in the $D$-theory it is useful to recall the structure of the corresponding
chiral primary field in the underlying conformal field theory and its corresponding
  Landau-Ginzburg model.
 It was pointed out by Boucher-Friedan-Kent \cite{bfk86}  that the chiral primary field
 $\vphi_\Om$ corresponding to the holomorphic form $\Om$ in an $N=2$
 superconformal field theory is characterized by its
 anomalous dimension $\Delta$ and U(1)-charge $Q$
  \beq
  (\Delta,Q)(\vphi_\Om) ~=~ \left(\frac{c}{6}, \frac{c}{3}\right), ~~~c=3\rmdim_\mathC~X.
  \eeq
 This translates into the same condition for the Landau-Ginzburg field
 $\Phi_\Om$ corresponding to $\vphi_\Om$.
  It follows from this and the central charge relation of the  tensor
  product (\ref{central-charge})
 that in the diagonal tensor model the field $\vphi_\Om$ corresponding to the
 holomorphic $\Om$-form is given by
 \beq
  \vphi_\Om ~=~ \prod_{i=0}^{n+1} \varphi_{k_i,-k_i,0}^{k_i,-k_i,0}.
  \eeq
  In the corresponding diagonal Landau-Ginzburg theory, determined by the superpotential
 \beq
  W(\Phi_i) ~=~ \sum_i \Phi_i^{k_i+2},
 \eeq
  the U(1) charges of the fields in the spectrum are given by
 \beq
   Q(\Phi_i^{\ell_i})~ =~ \frac{\ell_i}{k_i+2}.
  \eeq
  It follows from this that the field $\Phi_\Om$ must be constructed from the
highest powers of the fields $\Phi_i$ that is consistent with the ideal generated by
 the Landau-Ginzburg potential, i.e.
 \beq
 \Phi_\Om ~=~ \prod_i \Phi_i^{k_i},
 \eeq
 leading to
 \beq
 Q(\Phi_\Om) ~=~ \sum_i \frac{k_i}{k_i+2} =  \frac{c}{3}.
 \eeq
 The transition $k_A \longmapsto k_D$ is only possible for even $k$, therefore the highest power
  $\Phi^k$ is invariant under the  $\mathZ_2$ that defines the $D$-invariant, hence it survives the quotient.

\subsubsection{Part II: invariance of the Galois group}

It remains to show that while the Galois group generating the motivic orbit is different in general for
 diagonal and nondiagonal varieties  it is unchanged in the exchange of the affine invariants
  $k_A \longmapsto k_D$.
This follows by noting that
 $v_A = \rmlcm\{a,2\} = a$ since $a$ must be even: for $k_A$ one has the superpotential
 $W_A = \Phi^a + \Psi^2$, but the replacement $k_A \longmapsto k_D$
can be applied to even $k$.  The $D$-model superpotential $W_D=\tPhi^{a/2}+\tPhi\tPsi^2$ contributes
to the Galois field $K_v$
 $v_D = \rmlcm\left\{a/2,2\right\} = a$ since in general $a/2$ does not have to be even.
Combining these two facts shows that the motive-generating group is in fact the same for both
models.

\section{Applications}

\subsection{Automorphic motives from Kac-Moody algebras}
 One of the consequences of string theory is that it suggests a relation between
two-dimensional conformal field theory and the geometry of Calabi-Yau varieties.
 In the arithmetic framework this translates into
relations between $L$-functions arise in the context of a string theoretic
application of the arithmetic Langlands program in combination with Grothendieck's theory of motives.
The precise formulation of this framework involves
the interpretation of automorphic forms derived from pure or mixed motives that arise in Calabi-Yau
varieties to modular forms that come from Kac-Moody algebras. In this context one encounters results
like the following.

Let $X_2^{6\rmND}$ be a K3 surface of $D$-type in the weighted projective space $\mathP_{(1,1,2,2)}$
  and $X_2^{6\rmBP}$ the diagonal K3 surface in the space $\mathP_{(1,1,1,3)}$. The motivic $AD$-isomorphism
 then implies that the $L$-functions of the $\Om$-motives of these two K3 surfaces are identical.
 Because the latter hypersurface is known to be modular with an $\Om$-motivic modular form of
weight 3 and level 27 \cite{rs06} it follows that the $\Om$-motive of the nondiagonal surface is modular as well
 \beq
   f_\Om(X_2^{6\rmB,\rmND},q) ~=~ f_\Om(X_2^{6\rmA,\rmBP},q)  ~=~ f_{3,27}(q) ~\in ~ S_3(\G_0(27),e).
 \eeq
 More details can be found in \cite{rs13b}.

 The  $\Om$-motivic modular form $f_{3,27}(q)$ has physical significance  because
it admits an interpretation in terms of Hecke indefinite
modular forms that are associated to Kac-Peterson string functions,
 which are
 modular forms of half-integral weight \cite{kp84}. Such relations therefore link the spacetime geometry
 to the worldsheet theory $T_\Si$,  providing  a direct method to pass from the geometry of the extra
 dimensions to the worldsheet and vice versa.  In this way the automorphic approach provides a concrete
 realization of an emergent spacetime program in string theory.

\subsection{Arithmetic mirror symmetry}

One of the discoveries triggered by string theory is mirror symmetry \cite{cls89, gp89}.  As mentioned
in the introduction, from the zeta function point of view mirror symmetry has been puzzling since its
rational form treats the even and odd cohomology groups differently. It has thus been of some interest
to search for a quantum analog of the zeta function of varieties that would be invariant under mirror
symmetry.In the present discussion the focus instead is on the universal motivic $L$-function that
exists for any Calabi-Yau variety. In the class of Gepner models the construction of Greene and Plesser
 describes the mirror theory as a quotient of the Gepner model with respect to a discrete symmetry
 group. In some cases this group is $\mathZ_2$ and the fractional transform (\ref{fractional-transf})
 is a 1-1 transformation that maps the quotient model to hypersurface \cite{ls90}. Thus
 there exist pairs of mirror manifolds that are obtained by replacing the diagonal
affine invariant in the $A$-model by the affine $D$-invariant, leading to a weighted projective hypersurface
 of $D$-type. The motivic $AD$-isomorphism $H(M_\Om(X)) \cong H(M_\Om(s_D(X))$
  implies that the motivic $L$-functions of such mirror pairs
are identical $L_\Om(X,s) = L_\Om(s_D(X),s)$ independent of any input about the automorphic structure
 of the motives and their $L$-functions.

The motives that arise from exactly solvable mirror pairs among the Gepner models have quite high
rank because the degrees of the hypersurfaces are large. This makes an explicit analysis of the
automorphic structure of these models quite challenging. The lowest rank motive
 is obtained from mirror pairs of varieties of  degree 264. This arises from the diagonal
model $(1_A\otimes 6_A \otimes 31_A \otimes 86_A)_\rmGSO$, associated to the Calabi-Yau
 threefold embedded in the weighted projective space $\mathP_{(3,8,33,55,88,132)}$.
  The fractional  transform of the mirror quotient leads to the $D-$type hypersurface
 \beq
  \mathP_{(3,8,33,88,132)}[264]^{(57,81)}/\mathZ_2
  ~\cong ~ \mathP_{(3,8,66,88,99)}[264]^{(81,57)},
 \eeq
 leading  to a pure motive of rank 80. The other exactly solvable
mirror pairs constructed in \cite{ls89} lead to even higher rank motives.

 The virtue of the isomorphism established above is that even without a proof of automorphy
 it follows that the high-rank motives of exactly solvable mirror pairs in this class of
theories  have the same as the $L$-functions.
This is exactly as one would expect if the motives were automorphic and confirms this prediction of
 mirror symmetry, extending the elliptic curve results of \cite{rs07} to higher dimensions.

\section{A generalized map}

It is natural to ask whether there are other identities between $L$-functions that generalize
the $AD$-isomorphism above, independent of whether an exactly solvable model is known or not.

It turns out that this is the case.  Define the map
 \beq
   s_\rmND:~~ p_A ~=~ x^a+y^b ~~\lra ~~ x^{\frac{a+1}{2}} + xy^b, ~~~{\rm for~} a~{\rm odd}.
 \eeq
The central charge constraint  applied to the map $s_\rmND$
  \beq
   \frac{1}{a}+ \frac{1}{b} ~=~ \frac{2}{a+1} + \frac{1}{b}\left(1-\frac{2}{a+1}\right)
 \eeq
 leads to the constraint
  \beq
   a ~=~ \frac{b}{b-2}.
  \eeq
 Since $a$ is an integer the only solutions are $b=3,4$, leading to
  \beq
  (a,b) ~=~ (3,3), ~~~~(a,b)~=~ (2,4).
  \eeq
 The assumption that $a$ is odd then leaves just a single
 solution for the central charge constraint:
  \beq
  p_A ~=~ x^3+ y^3  ~~\longmapsto ~~ x^2+xy^3.
\eeq
This map is of interest because it explains $L$-function identities between nondiagonal and diagonal motives
 that are not based on exactly solvable models.

\vskip .4truein

{\bf Acknowledgement.} \hfill \break
 It is a pleasure to thank Monika Lynker for discussions. This research has been supported in part by a
 grant from the NSF under grant no. NSF-0969875.  The work reported here
  has benefitted from the support and hospitality of several institutions.
 Visits to the Werner Heisenberg Max Planck Institute in 2011 and to CERN in 2012 have greatly
  facilitated progress on this project and I'm grateful for support from the Max Planck
  Gesellschaft and CERN. Special thanks are due in particular to Dieter L\"ust in Munich and Wolfgang
 Lerche at CERN for making these visits possible and I thank the string theory groups at both
 institutions for their friendly hospitality.

\end{document}